\documentclass[12pt]{article}
\usepackage{amsmath,amsthm}
\usepackage{amsfonts}
\usepackage{amssymb}

\theoremstyle{definition}
\newtheorem{definition}{Definition}
\newtheorem{corollary}[definition]{Corollary}

\theoremstyle{plain}

\newtheorem{lemma}[definition]{Lemma}
\newtheorem{proposition}[definition]{Proposition}
\newtheorem{conjecture}[definition]{Conjecture}

\theoremstyle{remark}
\newtheorem{remark}{\sc Remark}

\title{\textbf{NONINVERTIBILITY, SEMISUPERMANIFOLDS AND CATEGORIES
REGULARIZATION\footnote{Invited talk given at the NATO Advanced Research Workshop
{\bf ``NONCOMMUTATIVE STRUCTURES IN MATHEMATICS AND PHYSICS''}
held in Kiev, 24-28 September 2000.}
}}
\author{\textbf{Steven Duplij}\thanks{E-mail:
Steven.A.Duplij@univer.kharkov.ua} \thanks{Internet:
http://gluon.physik.uni-kl.de/\~{}duplij}\\
\textit{Kharkov National University}\\
\textit{Kharkov 61001, Ukraine}
\\   and\\
\textbf{W\l adys\l aw Marcinek}\thanks{E-mail: wmar@ift.univ.wroc.pl}\\
\textit{Institute of
Theoretical Physics, University of Wroc\l aw}\\
\textit{Pl. Maxa Borna 9,
50-204 Wroc{\l}aw, Poland}
}

\date{21 December 2000}

\begin{document}
\maketitle

\begin{abstract}
The categories with noninvertible morphisms are studied analogously
to the semisupermanifolds with noninvertible transition functions.
The concepts of regular
$n$-cycles, obstruction and the regularization procedure are introduced
and investigated. It
is shown that the regularization of a category with nonivertible morphisms and
obstruction form a 2-category. The generalization of functors, Yang-Baxter
equation, (co-) algebras, (co-) modules and
some related structures
to the regular case is given.

\end{abstract}
\newpage

\section{Introduction}

In the supermanifold noninvertible generalization approach \cite
{duplij,dup-hab,dup18} we study here the obstructed cocycle conditions in
the category theory framework and extend them to such structures as
categories, functors, (co-) algebras, (co-) modules etc. This approach is
connected with the higher regularity concept \cite{dup/mar1} and
reconsidering the role of identities \cite{dup/mar}. The introduced category
regularization together with obstruction form a $2$-category. Similar
abstract structure generalizations were considered in topological QFT \cite
{bae/dol,cra/yet}, for $n$-categories \cite{baez,bae/dol2,bae/dol1},
near-group categories \cite{sie,gre/ser} (with noninvertible elements) and
weak Hopf algebras \cite{boh/nil/szl,nil1} in which the counit does not
satisfy $\varepsilon \left( ab\right) =\varepsilon \left( a\right)
\varepsilon \left( b\right) $ or satisfy first order (in our classification)
regularity conditions \cite{fangli1,fangli3}. We first show how to deal with
noninvertibility in the supermanifold theory \cite{berezin,leites} and then
apply this approach to more general structures.

\section{Supermanifolds and semisupermanifolds}

In the supermanifold theory \cite{berezin,leites,bar/bru/her} the phenomenon
of noninvertibility obviously arises from odd nilpotent elements and zero
divisors of Grassmann algebras (also in the infinite dimensional case \cite
{iva2}). Despite the invertibility question is quite natural, the answer is
not so simple and in some cases can be nontrivial, e.g. in some
superalgebras one can introduce invertible analog of an odd symbol \cite
{lei/pei}, or construct elements without number part which are not nilpotent
even topologically \cite{pes7}. Several guesses concerning inner
noninvertibility inherent in the supermanifold theory were made before, e.g.
``...there may be no inverse projection\footnote[0]{%
number part.} at all'' \cite{pen}, ``...a general SRS needs not have a body%
\footnotemark[0]  '' \cite{cra/rab}, or ``...a body\footnotemark[0]  may not
even exist in the most extreme examples'' \cite{bry3}. It were also
considered pure odd supermanifolds \cite{rab00,rab11} which give an important
counterexample to the Coleman-Mandula theorem ``...and provides us with a new,
missed so far, version of the Poincar\'e supergroup'' \cite{lei-arw}, exotic
supermanifolds
with nilpotent even coordinates \cite{kon/sch} and supergravity with
noninvertible vierbein \cite{dra/gun/the}. Some problems with odd directions
and therefore connected with noninvertibility in either event are described
in \cite{bry2,cat/rei/teo}, and a perspective list of supermanifold problems
was stated by D. Leites in \cite{lei2}.

The patch definition of a supermanifold $\frak{M}_{0}$ in most cases differs
from the patch definition of an ordinary manifold \cite{kosinski,lang} by
``super-'' terminology only and is well-known \cite{dewitt}. Let $%
\mathrel{\mathop{\mathop{\textstyle\bigcup}}
\limits_{\alpha}}\left\{ U_{\alpha },\varphi _{\alpha }\right\} $ is an
atlas of a \textsl{supermanifold} $\frak{M}_{0}$, then its gluing transition
functions $\Phi _{\alpha \beta }=\varphi _{\alpha }\circ \varphi _{\beta
}^{-1}$ satisfy the cocycle conditions
\begin{equation}
\Phi _{\alpha \beta }^{-1}=\Phi _{\beta \alpha },\;\;\;\;\;\Phi _{\alpha
\beta }\circ \Phi _{\beta \gamma }\circ \Phi _{\gamma \alpha }=1_{\alpha
\alpha }  \label{4}
\end{equation}

\noindent on overlaps $U_{\alpha }\cap U_{\beta }$ and on triple overlaps $%
U_{\alpha }\cap U_{\beta }\cap U_{\gamma }$ respectively, where $1_{\alpha
\alpha }\stackrel{\mathrm{def}}{=}id\,\left( U_{\alpha }\right) $. To obtain
a patch definition of an object analogous to supermanifold we try to weaken
demand of invertibility of coordinate maps $\varphi _{\alpha }$. Consider a
generalized superspace $\frak{M}$ covered by open sets $U_{\alpha }$ as $%
\frak{M}=\mathrel{\mathop{\bigcup}\limits_{\alpha}}U_{\alpha }$. We assume
here that the maps $\varphi _{\alpha }:U_{\alpha }\rightarrow V_{\alpha
}\subset \Bbb{R}^{n|m}$ are not all homeomorphisms, i.e. among them there
are noninvertible maps\footnote{%
Under $\Bbb{R}^{n|m}$ we imply some its noninvertible generalization \cite
{dup18}.}.

\begin{definition}
A\textsl{\ semisupermanifold }is a noninvertibly generalized superspace $%
\frak{M}$ represented as a semiatlas $\frak{M}=%
\mathrel{\mathop{\bigcup
}\limits_{\alpha}}\left\{ U_{\alpha },\varphi _{\alpha }\right\} $ with
invertible and noninvertible coordinate maps $\varphi _{\alpha }:U_{\alpha
}\rightarrow V_{\alpha }\subset \Bbb{R}^{n|m}$.
\end{definition}

We do not concretize here the details, how the invertibility appears here,
but instead we will describe it by some general relations between
semitransition functions and other objects. We The noninvertibly extended
gluing \textsl{semitransition functions }of a semisupermanifold are defined
by the equations
\begin{equation}
\Phi _{\alpha \beta }\circ \varphi _{\beta }=\varphi _{\alpha
},\;\;\;\;\;\;\Phi _{\beta \alpha }\circ \varphi _{\alpha }=\varphi _{\beta }
\label{6}
\end{equation}
instead of $\Phi _{\alpha \beta }=\varphi _{\alpha }\circ \varphi _{\beta
}^{-1}$, which obviously extends the class of funtions to noninvertible
ones. Then we assume that instead of (\ref{4}) the semitransition functions $%
\Phi _{\alpha \beta }$ of a semisupermanifold $\frak{M}$ satisfy the
following relations
\begin{equation}
\Phi _{\alpha \beta }\circ \Phi _{\beta \alpha }\circ \Phi _{\alpha \beta
}=\Phi _{\alpha \beta }  \label{8}
\end{equation}

\noindent on $U_{\alpha }\cap U_{\beta }$ overlaps (invertibility is
extended to regularity) and
\begin{eqnarray}
\Phi _{\alpha \beta }\circ \Phi _{\beta \gamma }\circ \Phi _{\gamma \alpha
}\circ \Phi _{\alpha \beta } &=&\Phi _{\alpha \beta },  \label{90} \\
\Phi _{\beta \gamma }\circ \Phi _{\gamma \alpha }\circ \Phi _{\alpha \beta
}\circ \Phi _{\beta \gamma } &=&\Phi _{\beta \gamma },  \label{91} \\
\Phi _{\gamma \alpha }\circ \Phi _{\alpha \beta }\circ \Phi _{\beta \gamma
}\circ \Phi _{\gamma \alpha } &=&\Phi _{\gamma \alpha }  \label{92}
\end{eqnarray}
on triple overlaps $U_{\alpha }\cap U_{\beta }\cap U_{\gamma }$ and
\begin{eqnarray}
\Phi _{\alpha \beta }\circ \Phi _{\beta \gamma }\circ \Phi _{\gamma \rho
}\circ \Phi _{\rho \alpha }\circ \Phi _{\alpha \beta } &=&\Phi _{\alpha
\beta },  \label{90w} \\
\Phi _{\beta \gamma }\circ \Phi _{\gamma \rho }\circ \Phi _{\rho \alpha
}\circ \Phi _{\alpha \beta }\circ \Phi _{\beta \gamma } &=&\Phi _{\beta
\gamma },  \label{91w} \\
\Phi _{\gamma \rho }\circ \Phi _{\rho \alpha }\circ \Phi _{\alpha \beta
}\circ \Phi _{\beta \gamma }\circ \Phi _{\gamma \rho } &=&\Phi _{\gamma \rho
},  \label{91v} \\
\Phi _{\rho \alpha }\circ \Phi _{\alpha \beta }\circ \Phi _{\beta \gamma
}\circ \Phi _{\gamma \rho }\circ \Phi _{\rho \alpha } &=&\Phi _{\rho \alpha }
\label{92w}
\end{eqnarray}
\noindent on $U_{\alpha }\cap U_{\beta }\cap U_{\gamma }\cap U_{\rho }$ . We
can write similar cycle relations to infinity and call them \textsl{tower
relations} which satisfy identically in the standard invertible case \textrm{
\cite{dewitt}}.

\begin{remark}
\label{rem}In any actions with noninvertible functions $\Phi _{\alpha \beta
} $ we are not allowed to cancel by them, because the semigroup of $\Phi
_{\alpha \beta }$'s is a semigroup without cancellation, and we are forced
to exploit the corresponding semigroup methods \cite{csa/thu,mag4}.
\end{remark}

\begin{conjecture}
The functions $\Phi _{\alpha \beta }$ satisfying the relations \textrm{(\ref
{8})--(\ref{92w})} can be viewed as some noninvertible generalization of the
transition functions as cocycles in the corresponding \v{C}ech cohomology of
coverings \textrm{\cite{maclane,switzer}}.
\end{conjecture}

\section{Obstructedness and additional orientation on semisupermanifolds}

The semisupermanifolds defined above belong to a class of so called
obstructed semisupermanifolds \cite{duplij,dup18} in the following sense.
Let us rewrite relations (\ref{4}) as the infinite series
\begin{equation}
\begin{array}{cc}
n=1: & \Phi _{\alpha \alpha }=1_{\alpha \alpha },
\end{array}
\label{11a}
\end{equation}
\begin{equation}
\begin{array}{cc}
n=2: & \Phi _{\alpha \beta }\circ \Phi _{\beta \alpha }=1_{\alpha \alpha },
\end{array}
\label{11b}
\end{equation}
\begin{equation}
\begin{array}{cc}
n=3: & \Phi _{\alpha \beta }\circ \Phi _{\beta \gamma }\circ \Phi _{\gamma
\alpha }=1_{\alpha \alpha },
\end{array}
\label{11c}
\end{equation}
\begin{equation}
\begin{array}{cc}
n=4: & \Phi _{\alpha \beta }\circ \Phi _{\beta \gamma }\circ \Phi _{\gamma
\delta }\circ \Phi _{\delta \alpha }=1_{\alpha \alpha }
\end{array}
\label{11d}
\end{equation}
\[
\begin{array}{cc}
\cdots & \cdots
\end{array}
\]

\begin{definition}
A semisupermanifold is called \textsl{obstructed},\textsl{\ }if some of the
cocycle conditions \textrm{(\ref{11a})--(\ref{11d})} are broken.
\end{definition}

It can happen that starting from some $n=n_{m}$ all higher cocycle
conditions hold valid.

\begin{definition}
\textsl{Obstructedness degree} of a semisupermanifold is a maximal $n_{m}$
for which the cocycle conditions \textrm{(\ref{11a})--(\ref{11d})} are
broken. If all of them hold valid, then $n_{m}\stackrel{\mathrm{def}}{=}0$.
\end{definition}

Obviously, that ordinary manifolds \cite{lang} \textrm{(}with invertible
transition functions\textrm{)} have vanishing obstructedness, and the
obstructedness degree for them is equal to zero, i.e. $n_{m}=0$.

\begin{remark}
The obstructed semisupermanifolds may have nonvanishing ordinary obstruction
which can be calculated extending the standard methods \textrm{\cite{berezin}%
} to the noninvertible case.
\end{remark}

Therefore, using the obstructedness degree $n_{m}$, we have possibility to
classify semisupermanifolds properly. Moreover, the pure soul supernumbers
do not contain unity. Obviously that obstructed semisupermanifolds cannot
have identity semitransition functions.

The orientation of ordinary manifolds is determined by the Jacobian sign of
transition functions $\Phi _{\alpha \beta }$ written in terms of local
coordinates on $U_{\alpha }\cap U_{\beta }$ overlaps \cite{kosinski,lang}.
Since this sign belong to $\Bbb{Z}_{2}$ , there exist two orientations on $%
U_{\alpha }$. Two overlapping charts are \textsl{consistently oriented} (or
\textsl{orientation preserving}) if $\Phi _{\alpha \beta }$ has positive
Jacobian, and a manifold is \textsl{orientable} if it can be covered by such
charts, thus there are two kinds of manifolds: orientable and nonorientable
\cite{lang}. In supersymmetric case the role of Jacobian plays Berezinian
\cite{berezin} which has a ``sign'' belonging to $\Bbb{Z}_{2}\oplus \Bbb{Z}%
_{2}$, and so there are four orientations on $U_{\alpha }$ and five
corresponding kinds of supermanifold orientability \cite{min,sha1}.

\begin{definition}
In case a nonvanishing Berezinian of $\Phi _{\alpha \beta }$ is nilpotent
\textrm{(}and so has no definite sign in the previous sense\textrm{)} there
exists additional \textsl{nilpotent orientation }on $U_{\alpha }$ of a
semisupermanifold.
\end{definition}

A degree of nilpotency of Berezinian allows us to classify
semisupermanifolds having nilpotent orientability (see e.g. \cite
{dup11,dup17}).

\section{Higher regularity and obstruction}

The above constructions have the general importance for \textit{any} set of
noninvertible mappings. The extension of $n=2$ cocycle given by (\ref{8})
can be viewed as some analogy with regular \cite{cli5} or pseudoinverse \cite
{mun/pen} elements in semigroups or generalized inverses in matrix theory
\cite{rao/mit}, category theory \cite{dav/rob} and theory of generalized
inverses of morphisms \cite{nashed}. The relations (\ref{90})--(\ref{92w})
and with other $n$ can be considered as noninvertible analogue of regularity
for higher cocycles. Therefore, by analogy with (\ref{8})--(\ref{92w}) it is
natural to formulate the general

\begin{definition}
An noninvertible mapping $\Phi _{\alpha \beta }$ is $n$\textsl{-regular}, if
it satisfies on overlaps $\stackrel{n}{\overbrace{U_{\alpha }\cap U_{\beta
}\cap \ldots \cap U_{\rho }}}$ to the following conditions
\begin{equation}
\stackrel{n+1}{\overbrace{\Phi _{\alpha \beta }\circ \Phi _{\beta \gamma
}\circ \ldots \circ \Phi _{\rho \alpha }\circ \Phi _{\alpha \beta }}}=\Phi
_{\alpha \beta }+\,perm.  \label{88}
\end{equation}
\end{definition}

The formula (\ref{8}) describes $3$-regular mappings, the relations (\ref{90}%
)--(\ref{92}) correspond to $4$-regular ones, and (\ref{90w})--(\ref{92w})
give $5$-regular mappings. Obviously that $3$-regularity coincides with the
ordinary regularity.

Let us consider a series of the selfmaps \textrm{$\mathbf{e}$}$_{\alpha
\alpha }^{\left( n\right) }:U_{\alpha }\rightarrow U_{\alpha }$ of a
semisupermanifold defined as
\begin{equation}
\mathrm{\mathbf{e}}_{\alpha \alpha }^{\left( 1\right) }=\Phi _{\alpha \alpha
},  \label{e11a}
\end{equation}
\begin{equation}
\mathrm{\mathbf{e}}_{\alpha \alpha }^{\left( 2\right) }=\Phi _{\alpha \beta
}\circ \Phi _{\beta \alpha },  \label{e11b}
\end{equation}
\begin{equation}
\mathrm{\mathbf{e}}_{\alpha \alpha }^{\left( 3\right) }=\Phi _{\alpha \beta
}\circ \Phi _{\beta \gamma }\circ \Phi _{\gamma \alpha },  \label{e11c}
\end{equation}
\begin{equation}
\mathrm{\mathbf{e}}_{\alpha \alpha }^{\left( 4\right) }=\Phi _{\alpha \beta
}\circ \Phi _{\beta \gamma }\circ \Phi _{\gamma \delta }\circ \Phi _{\delta
\alpha }  \label{e11d}
\end{equation}
\[
\begin{array}{cc}
\cdots & \cdots
\end{array}
\]

We will call \textrm{$\mathbf{e}$}$_{\alpha \alpha }^{\left( n\right) }$'s
\textsl{tower identities \textrm{(}or obstruction of }$U_{\alpha }$\textrm{)}%
. From (\ref{11a})--(\ref{11d}) it follows that for ordinary supermanifolds
obstruction coincide with the usual identity map
\begin{equation}
\mathrm{\mathbf{e}}_{\alpha \alpha }^{\left( n\right) ,ordinary}=1_{\alpha
\alpha }.  \label{e1}
\end{equation}

So the obstructedness degree can be treated as a maximal $n=n_{m}$ for which
tower identities differ from the identity, i.e. (\ref{e1}) is broken. The
obstruction gives the numerical measure of distinction of a
semisupermanifold from an ordinary supermanifold. When morphisms are
noninvertible (a semisupermanifold has a nonvanishing obstructedness), we
cannot ``return to the same point'', because in general \textrm{$\mathbf{e}$}%
$_{\alpha \alpha }^{\left( n\right) }\neq 1_{\alpha \alpha },$and we have to
consider ``nonclosed'' diagrams due to the fact that the relation \textrm{$%
\mathbf{e}$}$_{\alpha \alpha }^{\left( n\right) }\circ \Phi _{\alpha \beta
}=\Phi _{\alpha \beta }$ is noncancellative now (see \textsc{Remark}~\ref
{rem}).

Summarizing the above statements we propose the following intuitively
consistent changing of the standard diagram technique as applied to
noninvertible morphisms. In every case we get a new arrow which corresponds
to the additional multiplier, and so for $n=2$ we obtain \vskip15pt

\begin{center}
\setlength{\unitlength}{.3in}
\begin{picture}(15,3.5)
\put(4.5,2.8){\makebox(1,1){\small{\sf Invertible morphisms}}}
\put(4.6,2.2){{ ${\Phi}_{\alpha\beta}$}}
\put(4.6,0.7){{ ${\Phi}_{\beta\alpha}$}}
\put(4,1.8){\vector(1,0){2.1}}
\put(6.1,1.4){\vector(-1,0){2.1}}
\put(8,1.5){$\Longrightarrow$}
\put(10.5,2.8){\makebox(1,1){\small{\sf Noninvertible morphisms}}}
\put(10.6,0.6){{ ${\Phi}_{\beta\alpha}$}}
\put(10.6,2.2){{ ${\Phi}_{\alpha\beta}$}}
\put(10,1.9){\vector(1,0){2.1}}
\put(12.1,1.3){\vector(-1,0){2.1}}
\put(10,1.6){\vector(1,0){2.1}}
\put(0,1.5){\large{\sf n=2}}
\end{picture}
\end{center}

\noindent which describes the transition from (\ref{11b}) to (\ref{8}) and
presents the ordinary regularity condition for morphisms \cite
{dav/rob,nashed}. The most intriguing semicommutative diagram is the
triangle one

\begin{center}
\setlength{\unitlength}{.3in}
\begin{picture}(15,5)
\put(4.5,0){\makebox(1,1){\small{\sf Invertible morphisms}}}
\put(10.5,0){\makebox(1,1){\small{\sf Noninvertible morphisms}}}
\put(5.8,1.5){\vector(-1,1){1.68}}
\put(4.6,4.2){{ ${\Phi}_{\alpha\beta}$}}
\put(3.5,2){{ ${\Phi}_{\gamma\alpha}$}}
\put(4,3.5){\vector(1,0){2.1}}
\put(6.1,3.2){\vector(0,-1){2}}
\put(8,2.5){$\Longrightarrow$}
\put(13.3,2.5){\small+ {\it perm.}}
\put(6.5,2){{ ${\Phi}_{\beta\gamma}$}}
\put(9.5,2){{ ${\Phi}_{\gamma\alpha}$}}
\put(11.8,1.5){\vector(-1,1){1.68}}
\put(10.6,4.2){{ ${\Phi}_{\alpha\beta}$}}
\put(10,3.8){\vector(1,0){2.1}}
\put(10,3.4){\vector(1,0){2.1}}
\put(12.1,3.2){\vector(0,-1){2}}
\put(12.5,2){{ ${\Phi}_{\beta\gamma}$}}
\put(0,2.5){\large{\sf n=3}}
\end{picture}
\end{center}

\noindent which generalizes the cocycle condition (\ref{4}).

The higher $n$-regular semicommutative diagrams can be considered in the
framework of generalized categories \cite{bae/dol2,gre/ser,bre} in the
following way.

\section{Categories and $2$-categories}

There is an algebraic approach to the formalism considered in previous
sections based on the category theory \cite{dup/mar,dup/mar1}. A category $%
\mathcal{C}$ contains a collection $\mathcal{C}_{0}$ of objects and a
collection $\hom \left( \mathcal{C}\right) $ of arrows (morphisms) (see e.g.
\cite{maclane1}). The collection $\hom \left( \mathcal{C}\right) $ is the
union of mutually disjoint sets $\hom _{\mathcal{C}}(X,Y)$ of arrows $X%
\stackrel{f}{\longrightarrow }Y$ from $X$ to $Y$ defined for every pair of
objects $X,Y\in \mathcal{C}$. It may happen that for a pair $X,Y\in \mathcal{%
C}$ the set $\hom _{\mathcal{C}}(X,Y)$ is empty. The associative composition
of morphisms is also defined. By an equivalence in $\mathcal{C}$ we mean a
class of morphisms $\hom ^{\prime }(\mathcal{C})=\bigcup_{X,Y\in (\mathcal{C}%
_{0})}\hom _{\mathcal{C}}^{\prime }$ $(X,Y)$ where $\hom _{\mathcal{C}%
}^{\prime }(X,Y)$ is a subset of $\hom _{\mathcal{C}}(X,Y)$. Two objects $%
X,Y $ of the category $\mathcal{C}$ is equivalent if and only if there is an
morphism $X\stackrel{s}{\longrightarrow }Y$ in $\hom _{\mathcal{C}}^{\prime
}(X,Y)$ such that
\begin{equation}
\begin{array}{cc}
s^{-1}\circ s=id_{X}, & s\circ s^{-1}=id_{Y}
\end{array}
\end{equation}

Let $X=(X_{1},\cdots ,X_{n})$ be a sequence of objects of $\mathcal{C}$. Our
category can contains a class of \textit{noninvertible} morphisms \cite
{dav/rob,dup/mar1}. A (strict) $2$-category $\mathcal{C}$ consists of a
collection $\mathcal{C}_{0}$ of objects as $0$-cells and two collections of
morphisms: $\mathcal{C}_{1}$ and $\mathcal{C}_{2}$ called $1$-cells and $2$%
-cells, respectively \cite{bae/neu}. For every pair of objects $X,Y\in
\mathcal{C}_{0}$ there is a category $\mathcal{C}(X,Y)$ whose objects are $1$%
-cell $f:X\rightarrow Y$ in $\mathcal{C}_{1}$ and whose morphisms are $2$%
-cells. For a pair of $1$-cells $f,g\in \mathcal{C}_{1}$ there is a $2$-cell
$s:f\rightarrow g$ in $\mathcal{C}_{2}$. For every three objects $X,Y,Z\in
\mathcal{C}_{0}$ there is a bifunctor
\begin{equation}
c:\{\mathcal{C}(X,Y)\times \mathcal{C}(Y,Z)\longrightarrow \mathcal{C}(X,Z)\}
\end{equation}
which is called a composition of $1$-cells. There is an identity $1$-cell $%
id_{X}\in \mathcal{C}(X,X)$ which acts trivially on $\mathcal{C}(X,Y)$ or $%
\mathcal{C}(Y,X)$. There is also $2$-cell $id_{id_{X}}$ which acts trivially
on $2$-cells.

Let $\mathcal{C}$ be a category with equivalence. Then one can see that
collection of all equivalence classes of objects of $\mathcal{C}$ forms a $2$%
-category $\mathbf{C}(\mathcal{C})$. These classes are $0$-cells of $\mathbf{%
C}(\mathcal{C})$, 1-cells are classes of morphisms of $\mathcal{C}$. and $2$%
-cells are maps between these classes. Observe that $1$-cells of $\mathbf{C}(%
\mathcal{C})$ can be represented by morphisms of the underlying category $%
\mathcal{C}$, but such representation is not unique. One equivalence class
can be represented by several equivalent morphisms. One can define
2-morphisms on equivalence classes, and $\mathbf{C}(\mathcal{C})$ becomes a
2-category. If the category $\mathcal{C}$ is equipped with certain
additional structures, then one can transform them into $\mathbf{C}(\mathcal{%
C})$. If for instance $\mathcal{C}$ is monoidal category with product $%
\otimes :\mathcal{C}\times \mathcal{C}\longrightarrow \mathcal{C}$, then $%
\mathbf{C}(\mathcal{C})$ becomes the so-called semistrict monoidal $2$%
-category. This means that the product $\otimes $ (under some natural
conditions) is defined for all cells of the $2$-category $\mathbf{C}(%
\mathcal{C})$. In the case of braided categories one can obtain the
semistrict braided monoidal category \cite{bae/neu}. Algebras, coalgebras,
modules and comodules can be also included in this procedure. We apply such
method to regularize categories with noninvertible morphisms and obstruction
\cite{dup/mar,dup/mar1}.

\section{Categories and regularization}

Let $\mathcal{C}$ be a category with invertible and noninvertible morphisms
\cite{dup/mar} and equivalence. The equivalence in $\mathcal{C}$ is here
defined as the class of invertible morphisms in the category $\mathcal{C}$.

\begin{definition}
A sequence of morphisms
\begin{equation}
\begin{array}{c}
X_{1}\stackrel{f_{1}}{\longrightarrow }X_{2}\stackrel{f_{2}}{\longrightarrow
}\cdots \stackrel{f_{n-1}}{\longrightarrow }X_{n}\stackrel{f_{n}}{%
\longrightarrow }X_{1}\label{regu}
\end{array}
\end{equation}
such that there is an (endo-)morphism $\mathbf{e}_{X_{1}}^{\left( 3\right)
}:X_{1}\longrightarrow X_{1}$ defined uniquely by the following equation
\begin{equation}
\begin{array}{c}
\mathbf{e}_{X_{1}}^{\left( n\right) }:=f_{n}\circ \cdots \circ f_{2}\circ
f_{1}\label{egu}
\end{array}
\end{equation}
and subjects to the relation $f_{1}\circ f_{n}\circ \cdots \circ f_{2}\circ
f_{1}=f_{1}$ is said to be a \textit{regular }$n$\textit{-cycle} on $%
\mathcal{C}$ and it is denoted by $f=(f_{1},\ldots f_{n})$.
\end{definition}

The (endo-)morphisms $\mathbf{e}_{X_{i}}^{\left( n\right)
}:X_{i}\longrightarrow X_{i}$ corresponding for $i=2,\ldots ,n$ are defined
by a suitable cyclic permutation of above sequence.

\begin{definition}
The morphism $\mathbf{e}_{X}^{(n)}$ is said to be an obstruction of $X$. The
mapping $\mathbf{e}^{(n)}:X\in \mathcal{C}_{0}\rightarrow \mathbf{e}%
_{X}^{(n)}\in \hom (X,X)$ is called a regular $n$-cycle obstruction
structure on $\mathcal{C}$.
\end{definition}

If
\[
X_{1}\stackrel{g_{1}}{\longrightarrow }X_{2}^{\prime }\stackrel{g_{2}}{%
\longrightarrow }\cdots \stackrel{g_{n-1}}{\longrightarrow }X_{n}^{\prime }%
\stackrel{g_{n}}{\longrightarrow }X_{1}
\]
is an another $n$-tuple of morphisms such that $\mathbf{e}_{X_{1}}^{\left(
n\right) }:g_{n}\circ \cdots \circ g_{2}\circ g_{1}$, then we assume that $%
X_{i}^{\prime }$ is equivalent to $X_{i}$, for $i=2,\ldots ,n$.

\begin{definition}
A map $s:f\Rightarrow g$ which sends the object $X_{i}$ into equivalent
object $X_{i}^{\prime }$ and morphism $f_{i}$ into $g_{i}$ is said to be
obstruction $n$-cycle equivalence.
\end{definition}

We have the diagram
\begin{equation}
\begin{array}{lccccr}
&  & X_{2}\stackrel{f_{2}}{\longrightarrow} \cdots\stackrel{f_{n-1}}{%
\longrightarrow}X_{n} &  &  &  \\
& \stackrel{f_{1}}{\nearrow} &  & \stackrel{f_{n}}{\searrow} &  &  \\
X_{1} &  & \Downarrow s &  & X_{1} &  \\
& \stackrel{g_{1}}{\searrow} &  & \stackrel{g_{n}}{\nearrow} &  &  \\
&  & X^{\prime}_{2}\stackrel{g_{2}}{\longrightarrow}\cdots\stackrel{g_{n-1}}{%
\longrightarrow}X^{\prime}_{n} &  &  &
\end{array}
\end{equation}

\begin{lemma}
\label{lemma1}There is a one to one correspondence between equivalence
classes of regular $n$-cycles and regular $n$-cycle obstruction structures.
\end{lemma}

If $f=(f_{1},\ldots f_{n})$ is a class of regular $n$-cycles, then there is
the corresponding regular $n$-cycle obstruction structure $\mathbf{e}:X\in
\mathcal{C}_{0}\rightarrow \mathbf{e}_{X}\in \hom (X,X)$ such that the
relation (\ref{egu}) holds true. Let $\mathbf{e}^{(n)}:X\in \mathcal{C}%
_{0}\rightarrow \mathbf{e}_{X}^{(n)}\in \hom (X,X)$ be a regular $n$-cycle
obstruction in $\mathcal{C}$.

\begin{definition}
A morphism $\alpha :X\longrightarrow Y$ of the category $\mathcal{C}$ such
that
\begin{equation}
\alpha \circ \mathbf{e}_{X}^{\left( n\right) }=\mathbf{e}_{Y}^{(n)}\circ
\alpha
\end{equation}
is said to be a regular $n$-cycle obstruction morphism from $X$ to $Y$.
\end{definition}

It follows from (\ref{regu}) that the morphism $\alpha$ is in fact a
sequence of morphism $\alpha:= (\alpha_{1}, \ldots, \alpha_{n})$ such that
the diagram
\begin{equation}
\begin{array}{rccccccccl}
& X_{1} & \stackrel{f_{1}}{\longrightarrow} & X_{2} & \stackrel{f_{2}}{%
\longrightarrow}\cdots\stackrel{f_{n-1}}{\longrightarrow} & X_{n} &
\stackrel{f_{n}}{\longrightarrow} & X_{1} &  &  \\
\alpha_{1} & \downarrow &  & \downarrow &  & \downarrow &  & \downarrow &
\alpha_{1} &  \\
& Y_{1} & \stackrel{g_{1}}{\longrightarrow} & Y_{2} & \stackrel{g_{2}}{%
\longrightarrow}\cdots\stackrel{g_{n-1}}{\longrightarrow} & Y_{n} &
\stackrel{g_{n}}{\longrightarrow} & Y_{1} &  &
\end{array}
\end{equation}
is commutative.

\begin{definition}
A collection of all equivalence classes of objects $\mathcal{C}_{0}$ with
obstruction structures $\mathbf{e}^{(n)}:X\in \mathcal{C}_{0}\rightarrow
\mathbf{e}_{X}^{(n)}\in \hom (X,X)$ is denoted by $\Re eg_{n}(\mathcal{C})$
and called an obstruction $n$-cycle regularization of $\mathcal{C}$. The
class of all regular $n$-cycle morphisms from $X$ to $Y$ is denote by $\Re
eg_{n}(\mathcal{C})(X,Y)$.
\end{definition}

\begin{corollary}
It follows from the Lemma \ref{lemma1} that the map $s:\alpha
\longrightarrow \beta $ which sends an arbitrary regular $n$-cycle morphisms
$\alpha \in \Re eg_{n}(\mathcal{C})(X,X^{\prime })$ into a regular $n$-cycle
morphisms $\beta \in \Re eg_{n}(\mathcal{C})(X,X^{\prime })$ is a regular
obstruction $n$-cycle equivalence.
\end{corollary}

One can define $2$-morphisms and an associative composition of $2$-morphisms
such that $\Re eg_{n}(\mathcal{C})(X,Y)$ becomes a category for every two
objects $X,Y\in \mathcal{C}_{0}$. If $\alpha :X\longrightarrow Y$ and $\beta
:Y\longrightarrow Z$ are two $n$-cycle morphisms, then the composition $%
\beta \circ \alpha :X\rightarrow Z$ is also a $n$-cycle morphism. In this
way we obtain the composition as bifunctors
\begin{equation}
c^{\Re eg_{n}}:=\{\Re eg_{n}(\mathcal{C})(X,Y)\times \Re eg_{n}(\mathcal{C}%
)(Y,Z)\longrightarrow \Re eg_{n}(\mathcal{C})(X,Z)\}
\end{equation}
We summarize our considerations in the following lemma:

\begin{lemma}
The class $\Re eg_{n}(\mathcal{C})$ forms a (strict) $2$-category whose $0$%
-cells are equivalence classes of objects of $\mathcal{C}$ with
obstructions, whose $1$-cells are regular $n$-cycle obstruction morphisms,
and whose $2$-cells are regular obstruction $n$-cycle $2$-morphisms.
\end{lemma}

\section{Regularization of monoidal categories functors and Yang-Baxter
equation}

Let $\mathcal{C}=\mathcal{C}(I,\otimes )$ be a monoidal category, where $I$
is the unit object and $\otimes :\mathcal{C}\times \mathcal{C}%
\longrightarrow \mathcal{C}$ is the monoidal product \cite{yet,joy/str}. If
the following relation
\begin{equation}
\mathbf{e}_{X}^{(n)}\otimes \mathbf{e}_{Y}^{(n)}=\mathbf{e}_{X\otimes
Y}^{(n)}.
\end{equation}
holds true, then we have

\begin{proposition}
The monoidal product of two regular $n$-cycles $X_{1},\ldots ,X_{n}$ and $%
Y_{1},\ldots ,Y_{n}$ with obstruction $\mathbf{e}_{X_{1}}^{(n)}$, and $%
\mathbf{e}_{Y}^{(n)}$, respectively, is the regular $n$-cycle
\[
X_{1}\otimes Y_{1},\otimes \cdots \otimes X_{n}\otimes Y_{n}
\]
with the obstruction $\mathbf{e}_{X\otimes Y}^{(n)}$.
\end{proposition}

One can see that in this case $\Re eg_{n}(\mathcal{C})$ is the so-called
semistrict monoidal category \cite{bae/neu}.

Let $\mathcal{C}$ and $\mathcal{D}$ be two monoidal categories and let $\Re
eg_{n}(\mathcal{C}),\Re eg_{n}(\mathcal{D})$ be their regularization $2$%
-categories. We can introduce the notion of regular $2$-functors,
pseudonatural transformations and modifications. All definitions do not
changed, but the preservation of the identity can be replaced by the
requirement of preservation of obstruction morphisms $\mathbf{e}_{X}^{\left(
n\right) }$ and the invertibility is replaced by regularity. If, for
instance, there is a regular $2$-functor $\mathcal{F}:\Re eg_{n}(\mathcal{C}%
)\longrightarrow \Re eg_{n}(\mathcal{C})$, then in addition to the standard
definition \cite{maclane1} we have the following relation
\begin{equation}
\mathcal{F}(\mathbf{e}_{X})=\mathbf{e}_{\mathcal{F}(X)}.
\end{equation}

In the same manner we can ``regularize'' pseudo-natural transformations and
modifications \cite{bre}. Let $\Re eg_{n}(\mathcal{C})$ be a semistrict
monoidal $2$-category. A pseudo-natural transformations $B=\{B_{X,X^{\prime
}}:X\otimes X^{\prime }\rightarrow X^{\prime }\otimes X\}$ and two regular
modifications $B_{X\otimes Y,Z}$, $B_{X,Y\otimes Z}$ such that
\begin{equation}
\begin{array}{rcl}
& B_{X\otimes Y,Z} &  \\
X\otimes Y\otimes Z & \longrightarrow  & Y\otimes Z\otimes X \\
B_{X,Y}\otimes \mathbf{e}_{Z}\searrow  &  & \nearrow \mathbf{e}_{Y}\otimes
B_{X,Z} \\
& Y\otimes X\otimes Z &
\end{array}
\end{equation}
and
\begin{equation}
\begin{array}{rcl}
& B_{X,Y\otimes Z} &  \\
X\otimes Y\otimes Z & \longrightarrow  & Z\otimes X\otimes Y \\
\mathbf{e}_{X}\otimes B_{Y,Z}\searrow  &  & \nearrow B_{X,Z}\otimes \mathbf{e%
}_{Y} \\
& X\otimes Z\otimes Y &
\end{array}
\end{equation}
and
\begin{equation}
B_{X,X^{\prime }}\circ \mathbf{e}_{X\otimes X^{\prime }}=\mathbf{e}%
_{X^{\prime }\otimes X}\circ B_{X,X^{\prime }},
\end{equation}
are said to be a regular $n$-cycle braiding. Obviously, these operations
must satisfying all conditions of \cite{bae/neu} with two changes indicated
at the beginning of this section. Then the $2$-category $\Re eg_{n}(\mathcal{%
C})$ is called a semistrict regular $n$-cycle braided monoidal category.
This allows us to obtain here the following regular $n$-cycle Yang--Baxter
equation \cite{dup/mar,dup/mar1}
\begin{equation}
\mathbf{B}_{Y,Z,X}^{\left( 1\right) }\circ \mathbf{B}_{Y,X,Z}^{\left(
2\right) }\circ \mathbf{B}_{X,Y,Z}^{\left( 1\right) }=\mathbf{B}%
_{Z,X,Y}^{\left( 2\right) }\circ \mathbf{B}_{X,Z,Y}^{\left( 1\right) }\circ
\mathbf{B}_{X,Y,Z}^{\left( 2\right) },  \label{yb0}
\end{equation}
where the notation
\[
\mathbf{B}_{X,Y,Z}^{\left( 1\right) }=B_{X,Y}\otimes \mathbf{e}_{Z},\;\;%
\mathbf{B}_{X,Y,Z}^{\left( 2\right) }=\mathbf{e}_{X}\otimes B_{Y,Z}
\]
has been used and the obstruction $\mathbf{e}_{X}$ is exploited instead of
the identity $Id_{X}$. Solutions of the regular $n$-cycle Yang--Baxter
equation (\ref{yb0}) can be found by application of the endomorphism
semigroup methods used in \cite{fangli2,fangli3}.

\section{Regularization of algebras, coalgebras, modules and comodules}

Let $(\mathcal{C})$ be a monoidal category and $\Re eg_{n}(\mathcal{C})$ be
its regularization . It is known that an associative algebra in the category
$\mathcal{C}$ is an object $\mathcal{A}$ of this category such that there is
an associative multiplication $m:\mathcal{A}\otimes \mathcal{A}\rightarrow
\mathcal{A}$ which is also a morphism of this category. If the
multiplication is in addition a regular $n$-cycle morphism, then the algebra
$\mathcal{A}$ is said to be a regular $n$-cycle algebra. This means that we
have the relation
\begin{equation}
m\circ (\mathbf{e}_{\mathcal{A}}\otimes \mathbf{e}_{\mathcal{A}})=\mathbf{e}%
_{\mathcal{A}}\circ m.
\end{equation}

Obviously such multiplication not need to be unique. Denote by $\Re eg_{n}(%
\mathcal{C})(\mathcal{A}\otimes \mathcal{A},\mathcal{A})$ a class of all
such multiplications. We can see that a regular $n$-cycle $2$-morphisms $%
s:m\Rightarrow n$ which send the multiplication $m$ into a new one $n$
should be an algebra homomorphism. One can define regular $n$-cycle
coalgebra or bialgebra in a similar way. A comultiplication $\triangle :%
\mathcal{A}\longrightarrow \mathcal{A}\otimes \mathcal{A}$ can be
regularized according to the relation
\begin{equation}
\triangle \circ \mathbf{e}_{\mathcal{A}}=(\mathbf{e}_{\mathcal{A}}\otimes
\mathbf{e}_{\mathcal{A}})\circ \triangle .
\end{equation}

In this case we obtain a class $\Re eg_{n}(\mathcal{C})(\mathcal{A},\mathcal{%
A}\otimes \mathcal{A})$ of comultiplications.

Let $_{\mathcal{A}}\mathcal{C}$ be a category of all left $\mathcal{A}$
-modules, where $\mathcal{A}$ is a bialgebra. For the regularization $\Re
eg_{n}(\mathcal{_{\mathcal{A}}\mathcal{C})}$ of the $\mathcal{A}$--module
action $\rho _{M}:\mathcal{A}\otimes M\longrightarrow M$ we use the
following formula
\begin{equation}
\rho _{M}\circ (\mathbf{e}_{\mathcal{A}}\otimes \mathbf{e}_{M})=\mathbf{e}%
_{M}\circ \rho _{M},  \label{rmod}
\end{equation}
where $\rho _{M}:\mathcal{A}\otimes M\longrightarrow M$ is the left module
action of $\mathcal{A}$ on $M$. The class of all such module actions is
denoted by $\Re eg_{n}(_{\mathcal{A}}\mathcal{C})(\mathcal{A}\otimes
\mathcal{M},\mathcal{M})$. The monoidal operation in this category is given
as the following tensor product of $\mathcal{A}$-modules
\begin{equation}
\rho _{M\otimes N}:=(id_{M}\otimes \tau \otimes id_{N})\circ (\rho
_{M}\otimes \rho _{N})\circ (\triangle \otimes id_{M\otimes N}),
\end{equation}
where $\tau :\mathcal{A}\otimes M\rightarrow M\otimes \mathcal{A}$ is the
twist, i. e. $\tau (a\otimes m):=m\otimes a$ for every $a\in \mathcal{A}%
,m\in M$.

\begin{lemma}
For the tensor product of module actions we have the following formula
\begin{equation}
\rho _{M\otimes N}\circ (\mathbf{e}_{\mathcal{A}}\otimes \mathbf{e}%
_{M\otimes N})=\mathbf{e}_{M\otimes N}\circ \rho _{M\otimes N}.
\end{equation}
\end{lemma}

This lemma means that the tensor product of two module actions satisfy our
regularity condition if and only if these two actions also satisfy the
regularity condition (\ref{rmod}).

Observe that there is also a category $\mathcal{C}^{\mathcal{A}}$ of right $%
\mathcal{A}$-comodules, where $\mathcal{A}$ is an algebra. We can regularize
this category in the following way. For the coaction we have
\begin{equation}
\rho \circ \mathbf{e}_{\mathcal{A}}=(\mathbf{e}_{M}\otimes \mathbf{e}_{%
\mathcal{A}})\circ \rho _{M},
\end{equation}
and
\begin{equation}
\rho _{M\otimes N}:=(id_{M}\otimes m_{\mathcal{A}})\circ (id_{M}\otimes \tau
\otimes id_{N})\circ (\rho _{M}\otimes \rho _{N}),
\end{equation}
where $\tau :M\otimes N\rightarrow N\otimes M$ is the twist, $m_{\mathcal{A}%
}:\mathcal{A}\otimes \mathcal{A}\rightarrow \mathcal{A}$ is the
multiplication in $\mathcal{A}$.

\section*{Conclusions}

Thus noninvertible extension of many abstract structures can be done in
common general way: by introduction of the obstructions (or $n$-cycles) $%
\mathbf{e}$ which are analogs of units of the invertible case. In search of
possible analogies we observe that ``$\ln \mathbf{e}$'' can play the role of
first ``fundamental group'' for ``space'' of categories and vanishes for
invertible morphisms, while its difference from ``zero'' can be treated as
nontrivial ``noninvertible topology'' of such ``space''. We also note that
``nil-'' extension of supermanifolds -- semisupermanifolds \cite{dup14,dup18}
-- can be compared with the ``meta-'' extension of supermanifolds--
metamanifolds \cite{lei/ser1,lei/ser2} -- to find their complimentarity or
additivity and possibly for further generalizations simultaneously in both
ways.

\bigskip

\textbf{Acknowledgments}. One of the authors (S.D.) would like to thank
Andrzej Borowiec, Friedemann Brandt, Dimitry Leites, Jerzy Lukierski and
Volodymyr Lyubashenko for valuable discussions and Fang Li for fruitful
correspondence and rare reprints.


\end{document}